# Diluted Oxide Interfaces with Tunable Ground States

*Yulin Gan, Dennis Valbjørn Christensen, Yu Zhang, Hongrui Zhang, Krishnan Dileep, Zhicheng Zhong, Wei Niu, Damon James Carrad, Kion Norrman, Merlin von Soosten, Thomas sand Jespersen, Baogen Shen, Nicolas Gauquelin, Johan Verbeeck, Jirong Sun, Nini Pryds and Yunzhong Chen\**

Y. L. Gan, Dr. D. V. Christensen, Y. Zhang, Dr. W. Niu, Dr. K. Norrman, M. von Soosten, Prof. N. Pryds and Prof. Y. Z. Chen
Department of Energy Conversion and Storage, Technical University of Denmark, Risø Campus, 4000 Roskilde, Denmark
E-mail: yunc@dtu.dk

Y. Zhang, Dr. H. R. Zhang, Prof. B. G. Shen and Prof. J. R. Sun
National Laboratory for Condensed Matter Physics and Institute of Physics, Chinese Academy of Sciences, Beijing 100190, China

D. Krishnan, Dr. N. Gauquelin and Prof. J. Verbeeck
EMAT, University of Antwerp, Groenenborgerlaan 171, 2020 Antwerp, Belgium

Prof. Z. Zhong
Key Laboratory of Magnetic Materials and Devices, Ningbo Institute of Materials Technology and Engineering, Chinese Academy of Sciences, Ningbo, 315201, China

Dr. D. J. Carrad, M. von Soosten and Prof. T. S. Jespersen
Center for Quantum Devices, Niels Bohr Institute, University of Copenhagen, Universitetsparken 5, 2100 Copenhagen, Denmark

**Abstract:**

The metallic interface between two oxide insulators, such as LaAlO$_3$/SrTiO$_3$ (LAO/STO), provides new opportunities for electronics and spintronics. However, due to the presence of multiple orbital populations, tailoring the interfacial properties such as the ground state and metal-insulator transitions remains challenging. Here, we report an unforeseen tunability of the phase diagram of LAO/STO by alloying LAO with a ferromagnetic LaMnO$_3$ insulator without forming lattice disorder and at the same time without changing the polarity of the system. By increasing the Mn-doping level, $x$, of LaAl$_{1-x}$Mn$_x$O$_3$/STO ($0 \leq x \leq 1$), the interface undergoes a Lifshitz transition at $x = 0.225$ across a critical carrier density of $n_c = 2.8 \times 10^{13}$ cm$^{-2}$, where a peak $T_{SC} \approx 255$ mK of superconducting transition temperature is observed. Moreover, the LaAl$_{1-x}$Mn$_x$O$_3$ turns ferromagnetic at $x \geq 0.25$. Remarkably, at $x = 0.3$, where the metallic interface is populated by only $d_{xy}$ electrons and just before it becomes insulating, we achieve reproducibly a same device with both signatures of superconductivity and clear anomalous Hall effect ($7.6 \times 10^{12}$ cm$^{-2} < n_s \leq 1.1 \times 10^{13}$ cm$^{-2}$). This provides a unique and effective way to tailor oxide interfaces for designing on-demand electronic and spintronic devices.

The surface of $SrTiO_3$ (STO)[1] or its interface to another oxide insulator such as the polar perovskite $LaAlO_3$ (LAO)[2] or the spinel $\gamma$-$Al_2O_3$[3] can host a two-dimensional electron liquid (2DEL). In particular, the 2DEL at LAO/STO interface has been found to exhibit a wide spectrum of emergent phenomena such as 2D superconductivity (SC),[4-8] magnetism[9-11] and possible coexistence of these two,[12-14] providing an intriguing platform for multifunctional electronic devices.

The carriers in the 2DEL of STO-based heterostructures occupy the Ti $3d$ $t_{2g}$ orbitals, where the in-plane $d_{xy}$ bands often have a lower energy than the out-of-plane oriented $d_{xz}/d_{yz}$ due to the confinement of these orbitals at the interface.[1, 15] Among the systems investigated so far, the epitaxial polar/nonpolar LAO/STO interface has drawn intensive attention and shows a typical carrier density, $n_s$, of approximately $4 \times 10^{13}$ $cm^{-2}$.[4] This high carrier density results in the population of orbitals with both $d_{xy}$ and $d_{xz}/d_{yz}$ symmetry. The occupation of carriers in different subbands has previously been controlled reversibly using electrostatic gating, leading to a versatile tuning of the physical properties of the interface.[16] For example, by back gating the interface through the STO substrate, a Lifshitz transition was revealed at a critical carrier density of $n_c = 1.7 \times 10^{13}$ $cm^{-2}$,[6] above which the Fermi level starts to cross the bottom of the $d_{xz}/d_{yz}$ bands. In proximity to $n_c$, superconductivity exhibits a dome-shape as a function of gating voltage thus the carrier density.[5-8] Subsequent investigations reveal that the Lifshitz transition does not occur at a universal carrier density, but depends crucially on the electrostatic boundary conditions among other parameters.[17] Notably, electric-field-induced metal-insulator phase transitions have so far been largely limited to the critical film thickness before the conduction emerges.[18,19] Additionally, most back gating experiments show that the electron mobility is lowered with the decrease of carrier density,[20,21] although an opposite tendency, i.e. the electron mobility is increased upon decrease of carrier density, was recently observed in top gating configurations.[17] Finally, the electron mobility of LAO/STO interface at low temperatures is often below 1000 $cm^2V^{-1}s^{-1}$, and usually the interface turns non-metallic once the carrier density drops below $10^{13}\,cm^{-2}$.

The conventional way to create the 2DEL is to grow the LAO at a lower oxygen pressure below $10^{-4}$ mbar, and afterwards anneal the sample at a much higher oxygen pressure. Although this could lead to a high mobility sample (up to 6600 $cm^2V^{-1}s^{-1}$) with a low carrier density of the order of $10^{12}$ $cm^{-2}$,[22] this approach shows significant challenges in reproducing such extreme samples. The other way to create STO 2DEGs is to grow LAO at an oxygen pressure between $10^{-5}$-$10^{-3}$ mbar without post annealing. As expected, this might result in LAO/STO samples with relatively high content of oxygen vacancies as demonstrated by their resistance switching behavior.[23] Recent progress in surface and interface engineering of oxide heterostructures enables the realization of 2DELs with low carrier densities in the range of $10^{12}\,cm^{-2}$ while maintaining a high mobility above 10,000 $cm^2\,V^{-1}\,s^{-1}$ at 2 K.[24-27] This results in, for example, the observation of quantum Hall effect in delta-doped STO[27] and in a LaMnO$_3$ (LMO) buffered LAO/STO heterointerface.[24,28] Among the various capping or buffer layers of manganites and cuprates used,[26,29] the introduction of an electron sink of polar LMO as a buffer layer at the LAO/STO interface is of particular interest, which provides a more effective modulation-doping effect than the capping layer cases as well as a strong suppression of oxygen vacancies on the STO side.[24,30]

The parent compound LMO is a charge transfer insulator with the electronic configuration of Mn $3d\,t_{2g}^3 e_g^1$, where the $e_g$ level is split into $e_g^1$ and $e_g^2$ levels by the static Jahn-Teller distortion. Notably, the empty $e_g^2$ subband is often lower than the Ti $3d\,t_{2g}$ bands of STO where the Fermi level of the 2DEL resides.[31-34] In this vein, the presence of LMO (also for SrMnO$_3$)[35] in LAO/STO will result in the electrons transfer firstly to the Mn sites (forming $Mn^{2+}$) before the formation of itinerant electrons thus 2DEL in STO.[36] This provides an efficient pathway to suppressing the carrier density of the conducting interface. This principle applies regardless of whether LMO is inserted as a buffer layer at the interface (**Figure 1**a)[24,35] or is introduced as dopants into the STO side (Figure 1b)[37,38]. In the former case, the LaMnO$_3$ buffer layer can in principle trap electrons with a concentration of up to 1 e/uc. In the latter case, introducing a doping level of only 1-2.5% of Mn into STO can trap all the mobile electrons (approximately 1.6-4×10$^{13}$ cm$^{-2}$), but at the same time it introduces impurity scattering centers for the electron transport (also localized in STO). In both cases, very sharp metal to insulator transitions have been observed.[24,35,37,38] Still, a delicate control of the carrier density remains challenging. This has prevented a mapping of the evolution of the ground state and the electronic structure of LAO/STO as a function of the decrease of the carrier density across the metal-to-insulator transition. Here we propose a third possibility of designing an electron sink by doping LaMnO$_3$ into the LAO top layers (Figure 1c). Being able to control the Mn-doping level in the full range ($0 \leq x \leq 1$) of the resulting LaAl$_{1-x}$Mn$_x$O$_3$ (LAMO) films enables us to tailor the ground state of the 2DEL by using Mn as a knob to tune the carrier density deliberately. Compared to previous reports of diluting the LAO layer with nonpolar STO,[39] this new approach provides an opportunity to tune the carrier density of the interfacial 2DEL without changing the polar discontinuity across the interface (Supplementary Information, Figure S2a). Moreover, thick LaMnO$_3$ films are often ferromagnetic when epitaxially grown on STO.[40,41] Therefore, this approach can also be used to study the magnetic proximity effect at LAO/STO interfaces as reported in EuTiO$_3$-buffered LAO/STO.[42]

All LAMO ($0 \leq x \leq 1$) films are found to be epitaxially grown on the (001)-oriented TiO$_2$-terminated STO substrates with a layer-by-layer two-dimensional growth mode as monitored *in-situ* by intensity oscillations of reflective high-energy electron diffraction (RHEED) (Supplementary Information, Figure S1a). The epitaxial growth of LAMO on STO is also confirmed by X-ray diffraction (XRD) measurement (Supplementary Information, Figure S1b), where no signature of any impurity phases was detected for the whole doping range ($0 \leq x \leq 1$). Additionally, terrace surface of the LAMO surface with step heights of approximately 0.4 nm was also obtained by atomic force microscopy (AFM) measurements (Supplementary Information, Figure S1c). **Figure 2**a shows the typical high-angle annular dark-field (HAADF) scanning transmission electron microscopy (STEM) image for a LAMO/STO ($x = 0.3$) with film thickness, $t$, of 8 unit cells (uc). It reveals a coherent heterointerface without detectable defects or dislocations. Further investigations of the chemical composition by electron energy-loss spectroscopy (EELS) elemental mapping were performed. Figure 2b and c show the spatially-resolved Ti-$L_{2,3}$ edge and Mn-$L_{2,3}$ edge spectra, respectively. Figure 2e and f show the two dimensional EELS line scans on the evolution of the Ti-$L_{2,3}$, and Mn-$L_{2,3}$, respectively, across the LAMO layer. It indicates that Mn ions are confined almost exclusively within the LAMO layer with negligible down-ward diffusion into the STO substrate. The Ti ions outward diffusion is confined to 1 nm across the interface. Overall, the LAMO/STO interface turns out to be sharper than the LMO/STO interface.[43] The most remarkable features are the presence of $Mn^{2+}$ states at the LAMO top surface as well as the interface, as observed by a shift to the lower binding energy in the $L_3$ peak (from ≈ 642 eV to ≈ 640 eV). On the other hand, the central part of LAMO shows mainly the presence of $Mn^{3+}$. A profile of the spatial distribution of $Mn^{2+}$ and $Mn^{3+}$ across the interface is summarized in Figure 2g for which references of $Mn^{2+}$ and $Mn^{3+}$ have been fitted to the experimental spectra using reference spectra taken for MnO ($Mn^{2+}$) and $LaMnO_3$ ($Mn^{3+}$) (Supplementary Information, Figure S5).[44,45] It confirms that electrons are accumulated on the $Mn^{2+}$ sites in LAMO close to LAMO/STO interface. Notably, the $Mn^{2+}$ ions observed at the surface of LAMO could be attributed to surface structure degradation, such as La-deficiency,[30,40,46] which is a common phenomenon for manganite-based thin films.[30,46,47] The Ti-$L_{2,3}$ edge spectra show mainly contribution from $Ti^{4+}$, which means that the density of 2DEL on the STO side, if it exists, is rather low (beyond the detection limits of STEM-EELS). Shortly, the detection of $Mn^{2+}$ ions close to the LAMO/STO interface rather than $Ti^{3+}$ strongly suggests that the electrons of LAMO/STO (Supplementary Information, Figure S2) fill the interfacial Mn $e_g^2$ subbands before the Ti 3$d$ bands, a situation similar to the LMO-buffer layer case.[24] Therefore, the Mn dopants can serve as an electron sink.

In absence of Mn doping ($x = 0$), our LAO/STO interface shows a sharp transition from an insulating to a metallic state as the LAO layer reaches a critical thickness, $t = 4$ uc (**Figure 3**a), similar to previous reports.[48] The metallic LAO/STO is characterized by a typical sheet carrier density, $n_s = 1.1 \times 10^{14}$ cm$^{-2}$ at room temperature ($T = 300$ K). For low Mn-doping, the threshold thickness for the interface conduction is expected to be enhanced slightly (Supplementary Information, Figure S2). For $x = 0.2$, the critical thickness remains at $t = 4$ uc, although the sheet resistance, $R_s$, is higher than $x = 0$. For $x = 0.5$, the interfaces are always found to be highly insulating, independent of film thickness, $t$. Therefore, in the subsequent studies, we have focused on samples with $t = 8$ uc.

Figure 3b shows the temperature-dependent $R_s$ of LAMO (8 uc)/STO ($0 \leq x \leq 1$). Figure 3c and d summarize the room temperature $R_s$ and $n_s$ as a function of $x$. It shows that the LAMO (8 uc)/STO interface remains metallic at $x \leq 0.3$ with $n_s$ gradually decreasing from $1.1 \times 10^{14}$ cm$^{-2}$ at $x = 0$ to $5.9 \pm 1.3 \times 10^{12}$ cm$^{-2}$ at $x = 0.3$. When $x > 0.3$, the interface turns insulating. Notably, the transition at $x = 0.3$ is so sharp that minor changes in the carrier density result in dramatic changes in the ground state (as discussed in detail later). For example, one specific device among the 6 Hall-bar devices with $x = 0.3$ investigated, shows a temperature dependence of the sheet resistance displaying a sharp metal-insulator transition upon cooling (S6 in Figure 3b). Thus, $x = 0.3$ is determined to be the highest achievable doping level that leads to conducting interfaces in the LAMO/STO system. In conclusion, the transport measurement shows a significant reduction of the 2DEL carrier density at the LAMO/STO interface with increasing Mn doping, as expected for an electron sink effect.

Our in-depth magneto-transport measurements reveal dramatic changes in the electronic structure of the metallic LAMO/STO ($x \leq 0.3$) interfaces. As we show in the following, the conduction of these samples can be divided into three categories with respect to doping: (i) at $0 \leq x \leq 0.2$, the 2DEL hosts two or more kinds of carriers; (ii) at $0.225 \leq x \leq 0.275$, the 2DEL has only one type of carriers; and (iii) at $x = 0.3$, the 2DEL shows a single type of carriers, which shows anomalous Hall effect. **Figure 4**a-i show the Hall resistance ($R_{xy}$), the corresponding Hall coefficient ($R_H = R_{xy}/B$), and the magnetoresistance ($MR = R_{xx}(B)/R_{xx}(B=0)-1$), as a function of magnetic field ($B$) at 2 K for metallic LAMO/STO interfaces with different $x$. For $x \leq 0.2$, $R_{xy}$ curves are all nonlinear with respect to magnetic field, $B$ (Figure 4a and d), and the $MR$ traces follow a bell-like shape (Figure 4g), where the $MR$-$B$ curves display a U shape at low field, and shift to a bell-like shape at a characteristic magnetic field ($B_C$, $B_C = (n_1\mu_1+n_2\mu_2)/[(n_1+n_2)\mu_1\mu_2]$ in the two-band model as discussed later) that varies with each $x$. When plotting the Hall coefficient, $R_H$, as a function of $B$, a bell-like shape (Figure 4d) is also obtained with a kink occurring at almost the same $B_C$ as observed in $MR$ traces. These features suggest that the conductivity comes from two or more types of carriers.[6] By increasing $x$ to 0.225, the $R_{xy}$ curve behavior changes from nonlinear to linear (Figure 4b and e). This is the typical behavior for carrier conduction in a single band.[6] Meanwhile, the $MR$ trace shows a U shape with a small magnetoresistance, which is consistent with the Lorentz magnetoresistance of a single-band conductor where $MR \propto (\mu B)^2$ (Figure 4h).[49] Based on the intrinsic electronic structure of the STO 2DEL,[15] this strongly indicates that interface conduction comes from both $d_{xy}$ and $d_{xz}/d_{yz}$ orbitals at $x \leq 0.2$. Upon increasing the Mn-doping, the carrier density is suppressed and the Fermi surface of 2DEL at LAMO/STO interface experiences a Lifshitz transition that occurs when the $n_s$ is suppressed to a critical value of $n_c \approx 2.8 \times 10^{13}$ cm$^{-2}$ ($x = 0.225$), below which the conduction only comes from a single type of $d_{xy}$ carriers as reported previously.[6,17] The linear Hall effect persists up to $x = 0.275$ and its slope gradually increases as $x$ increases from $x = 0.225$ to $x = 0.275$, showing that the carrier density decreases as the Mn-doping increases. Upon further increase of the Mn doping to $x = 0.3$, the *MR* curve remains U-shaped (Figure 4c), which is consistent with the single band behavior expected below the Lifshitz transition point. The $R_{xy}$, however, turns nonlinear again with respect to the magnetic field (Figure 4c,f), but the shapes of $R_{xy}$ and $R_H$ are fundamentally different from the non-linearity observed at $x \leq 0.2$, especially the $R_H$ shows an opposite tendency in Figure 4d and f. These significant differences suggest that behavior of $R_{xy}$ and $R_H$ for $x = 0.3$ cannot be modelled by the typical behavior for multiband systems particularly at low magnetic fields (-4 T < $B$ < 4 T) and we therefore attribute the nonlinearity to the anomalous Hall effect without detectable hysteresis.[50] Here, a magnetization in the sample causes a Hall voltage to build up as a result of an asymmetric scattering of the 2DEL, which suggests that the interfacial 2DEL becomes spin-polarized or even a soft ferromagnetic. This is a possible situation since the magnetization measurement by superconducting quantum interference device (SQUID) of the samples shows signatures of ferromagnetism in the LAMO films as discussed later ($x \geq 0.25$, Supporting Information Figure S7), which can induce a magnetic ordering in the 2DEL by proximity effects. A similar ferromagnetism-related nonlinear Hall effect has been reported at the EuTiO$_3$-buffered LAO/STO with a much higher carrier density, where the conduction comes from multiple bands.[42] Therefore, our LAMO/STO ($x = 0.3$) might represent the first example that AHE is realized at metallic oxide interface based on STO with a single type of conducting carriers. Notably, the scalability of the anomalous Hall conductance with respect to the sample conductance (Supplementary Information, Figure S8d) indicates that the AHE of our samples results from an intrinsic mechanism rather than the skew-scattering mechanism.[50] Note that the possible origins of ferromagnetism by oxygen vacancies or magnetic impurities at LAO/STO interface are excluded here because of the strong suppression of oxygen vaccancies at $x=0.3$ and the linear Hall effect at $0.225 \leq x \leq 0.275$.

**Figure** 5a and b summarize the carrier density and electron mobility of the LAMO/STO interface at 2 K as a function of the Mn-doping, $x$. Note that for $x \leq 0.2$, a two-band model (see Supplementary Information, Figure S9) was used to extract the carrier densities and mobility for the $d_{xy}$ ($n_1$, $\mu_1$) and $d_{xz}/d_{yz}$ ($n_2$, $\mu_2$) carriers, where $n_s = n_1+n_2$. As shown in Figure 5a and b, when $x \leq 0.2$, the density ($n_1$) and mobility ($\mu_1$) of $d_{xy}$ band are roughly constant, with typical values of $n_1 \approx 4\times10^{13}$ cm$^{-2}$ and $\mu_1 \approx 200$ cm$^2$V$^{-1}$s$^{-1}$ at 2 K. In contrast, $n_2$ for $d_{xz}/d_{yz}$ electrons is about one order of magnitude lower than $n_1$, and experiences a decrease of 35% upon Mn-doping. $\mu_2$, which is much larger than $\mu_1$, shows a decrease of approximately 30% (from 2000 cm$^2$V$^{-1}$s$^{-1}$ to 1400 cm$^2$V$^{-1}$s$^{-1}$). These trends are similar to those revealed by field effect tuning,[17,20] which indicate that $d_{xz}/d_{yz}$ electrons with lower density often have higher mobility than the $d_{xy}$ electrons. This further confirms that Mn-doping influences primarily the $d_{xz}/d_{yz}$ electrons that are close to the Fermi level, as also confirmed by our density functional theory based tight-binding model calculations shown in Figure 5c.[36] At $x = 0.225$, the carrier density of the system is suppressed to the Lifshitz transition point of $n_1 = 2.8\times10^{13}$ cm$^{-2}$ and $n_2 = 0$, i.e. the 2DEL is only populated by $d_{xy}$ electrons for $x \geq 0.225$. Remarkably, a boost of $\mu_1$ with decrease of the carrier density starts once the $d_{xz}/d_{yz}$ bands are depopulated. As $\mu_1$ increases, we also observe an increase in the magnetoresistance (Figure 4h-i), as it is expected from the Lorentz magnetoresistance. It should be noted that at $x = 0.3$ ($n_s = n_1 = 2.9\times10^{12}$ cm$^{-2}$), where the presence of an AHE shows that the 2DEL interacts with magnetic moments, they still show a relatively high mobility of approximately 1900 cm$^2$V$^{-1}$s$^{-1}$ at 2K. At $x > 0.3$, the concentration of Mn as electron trap centers is so high that no electrons will transfer to the Ti 3$d$ $t_{2g}$ levels, so the interface turns out to be highly insulating. In our system, the critical density of Lifshitz transition is different from those in previous reports,[6,17] which implies that electrostatics, electronic correlations and doping levels are all contributing factors in determining the band structure of STO-based 2DELs.

It is usually reported that non-doped LAO/STO heterostructures have a 2D ground state that is superconducting.[4-8] Adding Mn dopants not only suppresses the carrier density as the conventional gating effect, but also introduces ferromagnetism in proximity to the interface when the Mn concentration is relatively high.[51] The temperature-dependent magnetization from the unpatterned samples of LAMO/STO heterostructures has been measured by SQUID magnetometers at temperatures down to 5 K (Supplementary Information, Figure S7). Note that these measurements primarily reflect the magnetic state of the LAMO top film. Although LMO/STO ($x = 1$, $t = 8$ uc) is ferromagnetic (FM), no magnetic signal is detected at $x \leq 0.225$. Nevertheless, paramagnetic or ferromagnetic-like behavior appears at $0.25 \leq x \leq 0.5$. This suggests that the AHE observed at interface of $x = 0.3$ could result from proximity-induced ferromagnetism of the $d_{xy}$ electrons.

To investigate superconductivity in the LAMO/STO system, the temperature dependence of the resistance was measured for all samples in a dilution refrigerator down to ≈ 15mK (**Figure** 6b). We find that for $0 \leq x < 0.3$, all samples become superconducting. Notably, our undoped LAO/STO ($x = 0$) has a $T_{SC}$ of 34 mK, much lower than other reports.[4,5] This is probably due to its high carrier density[8] and/or the relatively high content of oxygen vacancies on the STO side as the LAO/STO heterostructure is prepared without post-annealing. At $x = 0.3$, the samples are at the border between an insulating and metallic state, and they are therefore sensitive to even minor variations in the sample fabrication. By reproducing several Hall-bar devices (samples S1-S5) with $x = 0.3$ (on two different STO substrates), we find that devices S3-S5 with carrier densities $2.9 \times 10^{12}$ cm$^{-2}$ $\leq n_s \leq 0.76 \times 10^{13}$ cm$^{-2}$ (extracted at $T = 2$ K) did not show any sign of a superconducting transition. Samples S1 and S2 with $7.6 \times 10^{12}$ cm$^{-2}$ $< n_s \leq 1.1 \times 10^{13}$ cm$^{-2}$ cm$^{-2}$, showed a pronounced decrease for $T < 50$ mK, but did not reach a zero-resistance state at the lowest temperatures of the system. We interpret this decrease as the beginning of a transition to a superconducting state with a

$T_{SC}$ below our experimental base temperature (≈ 15 mK). The conventional definition of the superconducting transition temperature $R_s(T_{SC}) = 0.5 \times R_s(0.5 \text{ K})$ yields the result shown in Figure 6c, which summarizes the evolution of the superconducting temperature ($T_{SC}$) as a function of $n_s$. $T_{SC}$ reproduces a dome-like dependence on carrier density, as also reported before.[5-7] At $n_s \geq 2.8 \times 10^{13}$ cm$^{-2}$ where both $d_{xy}$ and $d_{xz}/d_{yz}$ conduction subbands are populated ($x \leq 0.2$), $T_{SC}$ displays an upward trend as the $d_{xz}/d_{yz}$ carrier density decreases. At the Lifshitz transition point, $n_c$, when the $d_{xz}/d_{yz}$ band is emptied, $T_{SC}$ approaches a peak value of $T_{SC}^{MAX} \approx 255$ mK and a further decrease in carrier density leads to a decrease of $T_{SC}$. Interestingly, devices with $7.6 \times 10^{12}$ cm$^{-2}$ < $n_s \leq 1.1 \times 10^{13}$ cm$^{-2}$ show both a magnetic LAMO top layer measured at $T \leq 21$ K with the SQUID magnetometer as well as a 2DEL that exhibits an AHE at 2 K and turns superconducting at low temperatures (0 mK ≤ $T_{SC} \leq 15$ mK). We note that we also observed a full superconducting transition where $R_s = 0$ is reached at $T_{SC} \approx 160$ mK in a van der Pauw sample ($n_s = 8.9 \times 10^{12}$ cm$^{-2}$) that also shows AHE (see Supplementary Information, **Figure** S11).

Different from the conventional gating technique[5-6] and the dilute doping of polar LAO with nonpolar STO,[35] we have demonstrated that a polar manganite electron sink can be employed to tune efficiently the electronic structure of LAO/STO heterointerface without changing the interface polarity. The consequent 2DEL experiences a Lifshitz transition at a critical density of $2.8 \times 10^{13}$ cm$^{-2}$, where a superconducting dome with a peak $T_{SC}$ of 255.7 mK is observed. At $n_s \leq 2.8 \times 10^{13}$ cm$^{-2}$, when the 2DEL is populated exclusively by $d_{xy}$ electrons, the electron mobility increases as $n_s$ decreases. In addition, at $n_s \leq 1.1 \times 10^{13}$ cm$^{-2}$, the 2DEL shows interactions with magnetic moments. Notably, at the lowest limit of $n_s = 2.9 \times 10^{12}$ cm$^{-2}$, the magnetic 2DEL still exhibits a relatively high electron mobility up to ≈ 1900 cm$^2$V$^{-1}$s$^{-1}$. This is different from the LAO/EuTiO$_3$/STO system,[42] where the magnetic EuO sublayer not only provides a magnetic proximity effect but also introduces strong magnetic scattering to the electrons located in the neighboring TiO$_2$ sublayer, leading to a low electron mobility.[42] In contrast, the high mobility 2DEL at the LAMO/STO ($x$ = 0.3) stays on the STO side, which is spatially separated from the magnetic (Al$_{1-x}$Mn$_x$)O$_2$ layer by the LaO sublayer, therefore they can still exhibit high mobility.

Moreover, at $x$ = 0.3 for $7.6 \times 10^{12}$ cm$^{-2}$ < $n_s$ ≤ $1.1 \times 10^{13}$ cm$^{-2}$, we observe SC, AHE and a magnetic LAMO top layer in the same device. In a conventional picture, superconductivity and ferromagnetism are mutually exclusive phenomena, and their coexistence at LAO/STO interface has been suggested to result from ferromagnetism and SC stemming from different bands that are spatially separated from each other.[52] However, here we observe AHE and superconductivity in our 2DEL with only population of $d_{xy}$ bands. It, therefore provides a unique and intriguing system, and calls for further work in understanding the coexistence of superconductivity and ferromagnetism at complex oxide interfaces.

**Acknowledgements**
We thank the technical help from J. Geyti. J. R. Sun acknowledges the support of the National Basic Research of China (2016YFA0300701), the National Natural Science Foundation of China (11520101002), and the Key Program of the Chinese Academy of Sciences. N. Gauquelin, D. Krishnan and J. Verbeeck acknowledge funding from the Geconcentreerde Onderzoekacties (GOA) project "Solarpaint" of the University of Antwerp, Belgium.

**Conflict of Interest**
The authors declare no conflict of interest.

**Keywords**
Oxide interfaces, two-dimensional electron liquid, metal-insulator transitions, superconductivity, anomalous Hall effect

**References**

[1]     A. F. Santander-Syro, F. Fortuna, C. Bareille, T. C. Rodel, G. Landolt, N. C. Plumb, J. H. Dil, M. Radovic, *Nat. Mater.* **2014**, *13*, 1085.

[2] A. Ohtomo, H. Y. Hwang, *Nature* **2004**, *427*, 423;

[3] Y. Z. Chen, N. Bovet, T. Kasama, W. W. Gao, S. Yazdi, C. Ma, N. Pryds, S. Linderoth, *Adv. Mater.* **2014**, *26*, 1462.

[4] N. Reyren, S. Thiel, A. D. Caviglia, L. Fitting Kourkoutis, G. Hammerl, C. Richter, C. W. Schneider, T. Kopp, A.-S. Rüetschi, D. Jaccard, M. Gabay, D. A. Muller, J. M. Triscone, J. Mannhart, *Science* **2007**, *317*, 1196.

[5] A. D. Caviglia, S. Gariglio, N. Reyren, D. Jaccard, T. Schneider, M. Gabay, S. Thiel, G. Hammerl, J. Mannhart, J. M. Triscone, *Nature* **2008**, *456*, 624.

[6] A. Joshua, S. Pecker, J. Ruhman, E. Altman, S. Ilani, *Nat. Commun.* **2012**, *3*, 1129.

[7] G. E. D. K. Prawiroatmodjo, F. Trier, D. V. Christensen, Y. Chen, N. Pryds, T. S. Jespersen, *Phys. Rev. B* **2016**, *93*, 184504.

[8] G. Herranz, G. Singh, N. Bergeal, A. Jouan, J. Lesueur, J. Gazquez, M. Varela, M. Scigaj, N. Dix, F. Sanchez, J. Fontcuberta, *Nat. Commun.* **2015**, *6*, 6028.

[9] A. Brinkman, M. Huijben, M. van Zalk, J. Huijben, U. Zeitler, J. C. Maan, W. G. van der Wiel, G. Rijnders, D. H. Blank, H. Hilgenkamp, *Nat. Mater.* **2007**, *6*, 493.

[10] B. Kalisky, J. A. Bert, B. B. Klopfer, C. Bell, H. K. Sato, M. Hosoda, Y. Hikita, H. Y. Hwang, K. A. Moler, *Nat. Commun.* **2012**, *3*, 922.

[11] F. Bi, M. Huang, S. Ryu, H. Lee, C. W. Bark, C. B. Eom, P. Irvin, J. Levy, *Nat. Commun.* **2014**, *5*, 5019.

[12] J. A. Bert, B. Kalisky, C. Bell, M. Kim, Y. Hikita, H. Y. Hwang, K. A. Moler, *Nat. Phys.* **2011**, *7*, 767.

[13] D. A. Dikin, M. Mehta, C. W. Bark, C. M. Folkman, C. B. Eom, V. Chandrasekhar, *Phys. Rev. Lett.* **2011**, *107*, 056802.

[14] L. Li, C. Richter, J. Mannhart, R. C. Ashoori, *Nat. Phys.* **2011**, *7*, 762.

[15] M. Gabay, J.-M. Triscone, *Nat. Phys.* **2013**, *9*, 610.

[16] J. Biscaras, N. Bergeal, S. Hurand, C. Grossetete, A. Rastogi, R. C. Budhani, D. LeBoeuf, C. Proust, J. Lesueur, *Phys. Rev. Lett.* **2012**, *108*, 247004.

[17] A. E. Smink, J. C. de Boer, M. P. Stehno, A. Brinkman, W. G. van der Wiel, H. Hilgenkamp, *Phys. Rev. Lett.* **2017**, *118*, 106401.

[18] L. Chen, J. Li, Y. Tang, Y. Y. Pai, Y. Chen, N. Pryds, P. Irvin, J. Levy, *Adv. Mater.* **2018**, 1801794.

[19] W.-N. Lin, J.-F. Ding, S.-X. Wu, Y.-F. Li, J. Lourembam, S. Shannigrahi, S.-J. Wang, T. Wu, *Adv. Mater. Interfaces* **2014**, *1*, 1300001.

[20] C. Bell, S. Harashima, Y. Kozuka, M. Kim, B. G. Kim, Y. Hikita, H. Y. Hwang, *Phys. Rev. Lett.* **2009**, *103*, 226802.

[21] W. Liu, S. Gariglio, A. Fête, D. Li, M. Boselli, D. Stornaiuolo, J. M. Triscone, *APL Mater.* **2015**, *3*, 062805.

[22] A. D. Caviglia, S. Gariglio, C. Cancellieri, B. Sacepe, A. Fete, N. Reyren, M. Gabay, A. F. Morpurgo, J. M. Triscone, *Phys. Rev. Lett.* **2010**, *105*, 236802.

[23] S. Wu, X. Luo, S. Turner, H. Peng, W. Lin, J. Ding, A. David, B. Wang, G. Van Tendeloo, J. Wang, T. Wu, *Phys. Rev. X* **2013**, *3*, 041027.

[24] Y. Z. Chen, F. Trier, T. Wijnands, R. J. Green, N. Gauquelin, R. Egoavil, D. V. Christensen, G. Koster, M. Huijben, N. Bovet, S. Macke, F. He, R. Sutarto, N. H. Andersen, J. A. Sulpizio, M. Honig, G. E. Prawiroatmodjo, T. S. Jespersen, S. Linderoth, S. Ilani, J. Verbeeck, G. Van Tendeloo, G. Rijnders, G. A. Sawatzky, N. Pryds, *Nat. Mater.* **2015**, *14*, 801.

[25] Y. Xie, C. Bell, Y. Hikita, S. Harashima, H. Y. Hwang, *Adv. Mater.* **2013**, *25*, 4735.

[26] M. Huijben, G. Koster, M. K. Kruize, S. Wenderich, J. Verbeeck, S. Bals, E. Slooten, B. Shi, H. J. A. Molegraaf, J. E. Kleibeuker, S. van Aert, J. B. Goedkoop, A. Brinkman, D. H. A.

Blank, M. S. Golden, G. van Tendeloo, H. Hilgenkamp, G. Rijnders, *Adv. Funct. Mater.* **2013**, *23*, 5240.

[27]     Y. Matsubara, K. S. Takahashi, M. S. Bahramy, Y. Kozuka, D. Maryenko, J. Falson, A. Tsukazaki, Y. Tokura, M. Kawasaki, *Nat. Commun.* **2016**, *7*, 11631.

[28]     F. Trier, G. E. Prawiroatmodjo, Z. Zhong, D. V. Christensen, M. von Soosten, A. Bhowmik, J. M. Lastra, Y. Chen, T. S. Jespersen, N. Pryds, *Phys. Rev. Lett.* **2016**, *117*, 096804.

[29]     Y. J. Shi, S. Wang, Y. Zhou, H. F. Ding, D. Wu, *Appl. Phys. Lett.* **2013**, *102*, 071605.

[30]     Y. Chen, R. J. Green, R. Sutarto, F. He, S. Linderoth, G. A. Sawatzky, N. Pryds, *Nano Lett.* **2017**, *17*, 7062.

[31]     A. Sawa, A. Yamamoto, H. Yamada, T. Fujii, M. Kawasaki, J. Matsuno, Y. Tokura, *Appl. Phys. Lett.* **2007**, *90*, 252102.

[32]     Y. Z. Chen, J. R. Sun, A. D. Wei, W. M. Lu, S. Liang, B. G. Shen, *Appl. Phys. Lett.* **2008**, *93*, 152515.

[33]     J. H. Jung, K. H. Kim, D. J. Eom, T. W. Noh, E. J. Choi, J. Yu, Y. S. Kwon, Y. Chung, *Phys. Rev. B* **1997**, *55*, 15 489.

[34]     X. Zhai, C. S. Mohapatra, A. B. Shah, J. M. Zuo, J. N. Eckstein, *Adv. Mater.* **2010**, *22*, 1136.

[35]     Y. Z. Chen, Y. L. Gan, D. V. Christensen, Y. Zhang, N. Pryds, *J. Appl. Phys.* **2017**, *121*, 095305.

[36]     Z. Zhong, P. Hansmann, *Phys. Rev. X* **2017**, *7*, 011023.

[37]     T. Fix, J. L. MacManus-Driscoll, M. G. Blamire, *Appl. Phys. Lett.* **2009**, *94*, 172101.

[38]     T. Fix, F. Schoofs, J. L. Macmanus-Driscoll, M. G. Blamire, *Phys. Rev. Lett.* **2009**, *103*, 166802.

[39] M. L. Reinle-Schmitt, C. Cancellieri, D. Li, D. Fontaine, M. Medarde, E. Pomjakushina, C. W. Schneider, S. Gariglio, P. Ghosez, J. M. Triscone, P. R. Willmott, *Nat. Commun.* **2012**, *3*, 932.

[40] Z. Chen, Z. Chen, Z. Q. Liu, M. E. Holtz, C. J. Li, X. R. Wang, W. M. Lu, M. Motapothula, L. S. Fan, J. A. Turcaud, L. R. Dedon, C. Frederick, R. J. Xu, R. Gao, A. T. N'Diaye, E. Arenholz, J. A. Mundy, T. Venkatesan, D. A. Muller, L. W. Wang, J. Liu, L. W. Martin, *Phys. Rev. Lett.* **2017**, *119*, 156801.

[41] W. Niu, W. Liu, M. Gu, Y. Chen, X. Zhang, M. Zhang, Y. Chen, J. Wang, J. Du, F. Song, X. Pan, N. Pryds, X. Wang, P. Wang, Y. Xu, Y. Chen, R. Zhang, *Adv. Electron. Mater.* **2018**, 1800055.

[42] D. Stornaiuolo, C. Cantoni, G. M. De Luca, R. Di Capua, E. Di Gennaro, G. Ghiringhelli, B. Jouault, D. Marre, D. Massarotti, F. Miletto Granozio, I. Pallecchi, C. Piamonteze, S. Rusponi, F. Tafuri, M. Salluzzo, *Nat. Mater.* **2016**, *15*, 278.

[43] J. A. Mundy, Y. Hikita, T. Hidaka, T. Yajima, T. Higuchi, H. Y. Hwang, D. A. Muller, L. F. Kourkoutis, *Nat. Commun.* **2014**, *5*, 3464.

[44] J. Verbeeck, S. Van Aert, G. Bertoni, *Ultramicroscopy* **2006**, *106*, 976.

[45] R. Egoavil, S. Huhn, M. Jungbauer, N. Gauquelin, A. Beche, G. Van Tendeloo, J. Verbeeck, V. Moshnyaga, *Nanoscale* **2015**, *7*, 9835.

[46] M. P. de Jong, I. Bergenti, V. A. Dediu, M. Fahlman, M. Marsi, C. Taliani, *Phys. Rev. B* **2005**, *71*, 014434.

[47] Z. Liao, N. Gauquelin, R. J. Green, S. Macke, J. Gonnissen, S. Thomas, Z. Zhong, L. Li, L. Si, S. Van Aert, P. Hansmann, K. Held, J. Xia, J. Verbeeck, G. Van Tendeloo, G. A. Sawatzky, G. Koster, M. Huijben, G. Rijnders, *Adv. Funct. Mater.* **2017**, *27*, 1606717.

[48]     S. Thiel, G. Hammerl, A. Schmehl, C. W. Schneider, J. Mannhart, *Science* **2006**, *313*, 1942.

[49]     K. Seeger, *Semiconductor Physics: an introduction*, Springer-Verlag Berlin Heidelberg, New York, **2004**.

[50]     N. Nagaosa, J. Sinova, S. Onoda, A. H. MacDonald, N. P. Ong, *Rev. Mod. Phys.* **2010**, *82*, 1539.

[51]     H. J. Liu, J. C. Lin, Y. W. Fang, J. C. Wang, B. C. Huang, X. Gao, R. Huang, P. R. Dean, P. D. Hatton, Y. Y. Chin, H. J. Lin, C. T. Chen, Y. Ikuhara, Y. P. Chiu, C. S. Chang, C. G. Duan, Q. He, Y. H. Chu, *Adv. Mater.* **2016**, *28*, 9142.

[52]     S. Banerjee, O. Erten, M. Randeria, *Nat. Phys.* **2013**, *9*, 626.

**Figures**

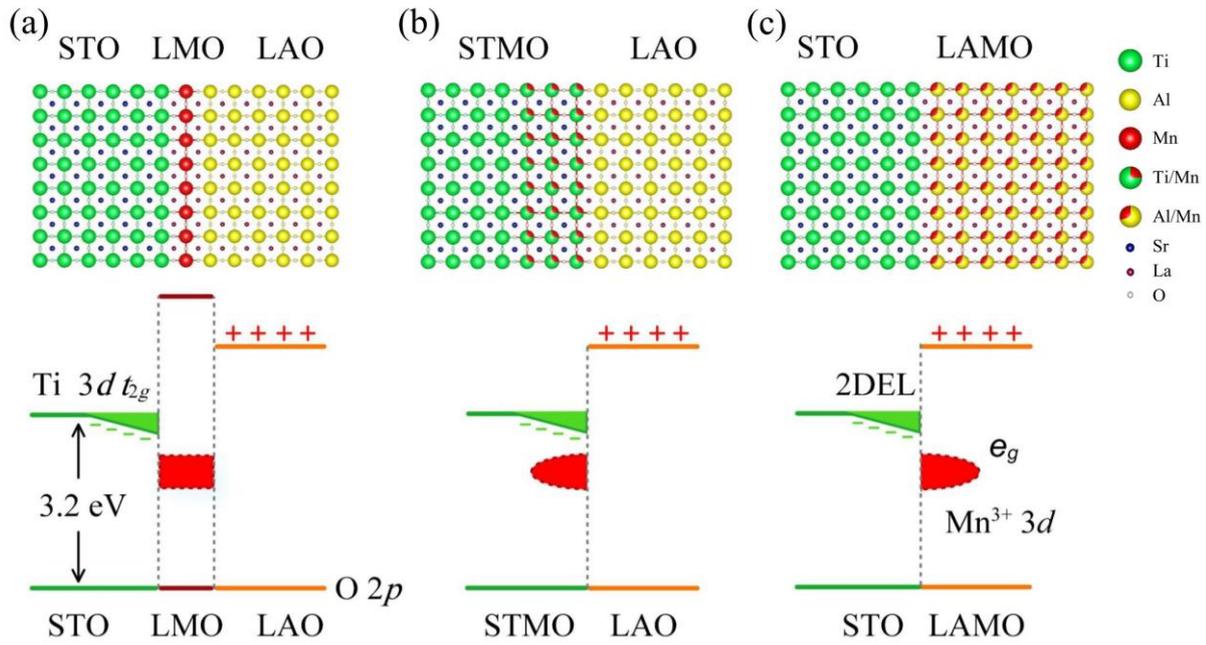

**Figure 1.** Three ways to introduce an electron sink at the LAO/STO interface and the corresponding band alignment: for the case of Mn dopants, a) buffer layer configuration of polar LaMnO$_3$ inserted between LAO and STO; b) small Mn-doping on the nonpolar STO side; c) Mn-dopants on the polar LAO side.

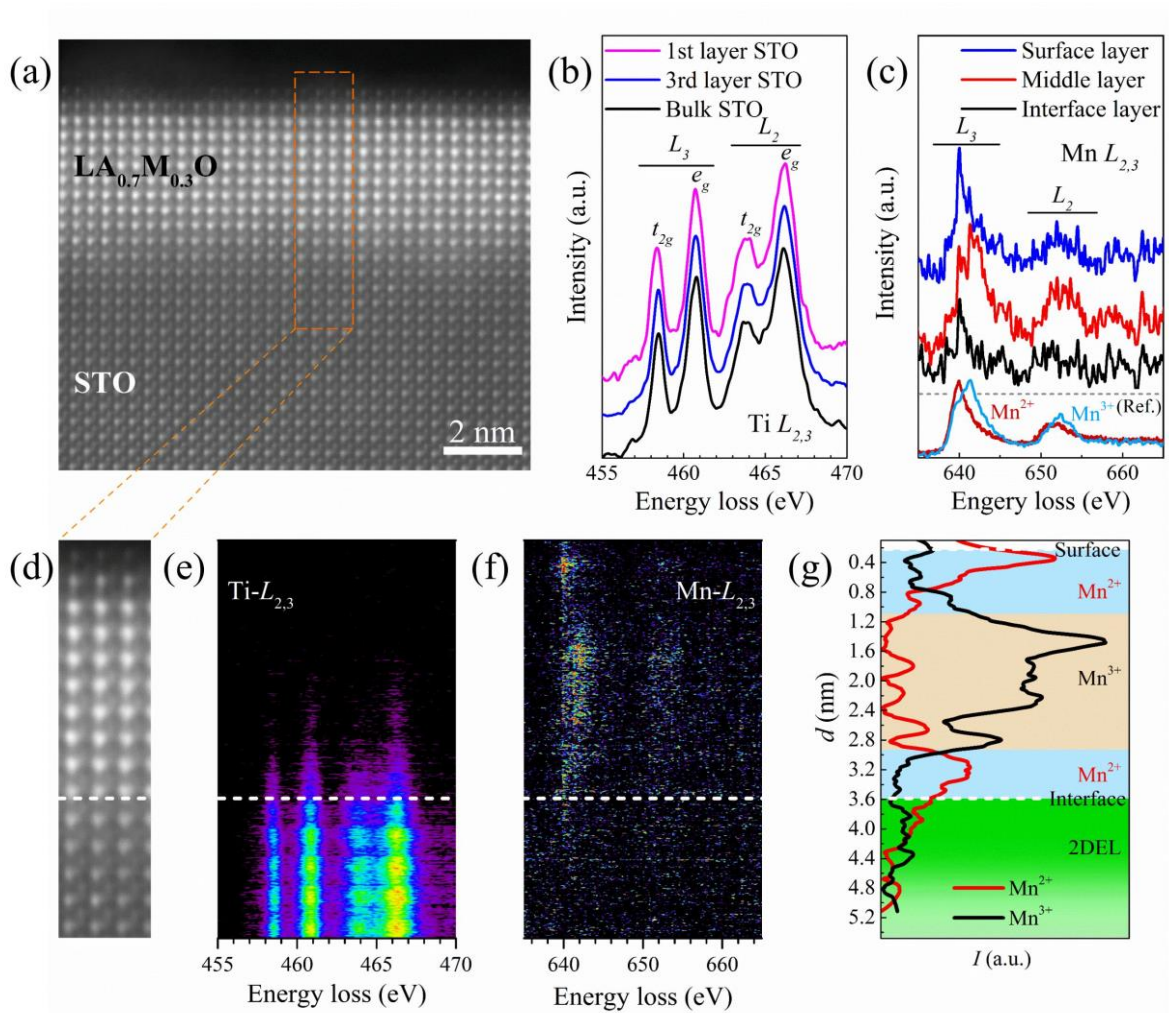

**Figure 2.** Sharp interface and electronic reconstruction of the epitaxial LAMO/STO heterostucture. a) HAADF-STEM images of an 8 uc LAMO ($x = 0.3$)/STO heterostructure. The brighter LaO layers determine the LAMO layer. b) The Ti $L_{2,3}$ energy loss near edge structure (ELNES) spectra from bulk STO, 3rd layer from the interface and the 1st layer before the interface. c) The Mn $L_{2,3}$ ELNES spectra integrated over the first 2~3 unit cells from the interface (interface layer), the next 5~6 unit cells from the interface (middle layer) and the last 2 unit cells near the surface (surface layer). The two standard spectra used for the fitting are from reference materials MnO ($Mn^{2+}$) and LaMnO$_3$ ($Mn^{3+}$) (ref. 44,45). e) and f) Spatially resolved Ti $L_{2,3}$ and Mn $L_{2,3}$ EELS spectra, respectively, in the selected area (d). g) The profile of $Mn^{2+}$ and $Mn^{3+}$ across the film.

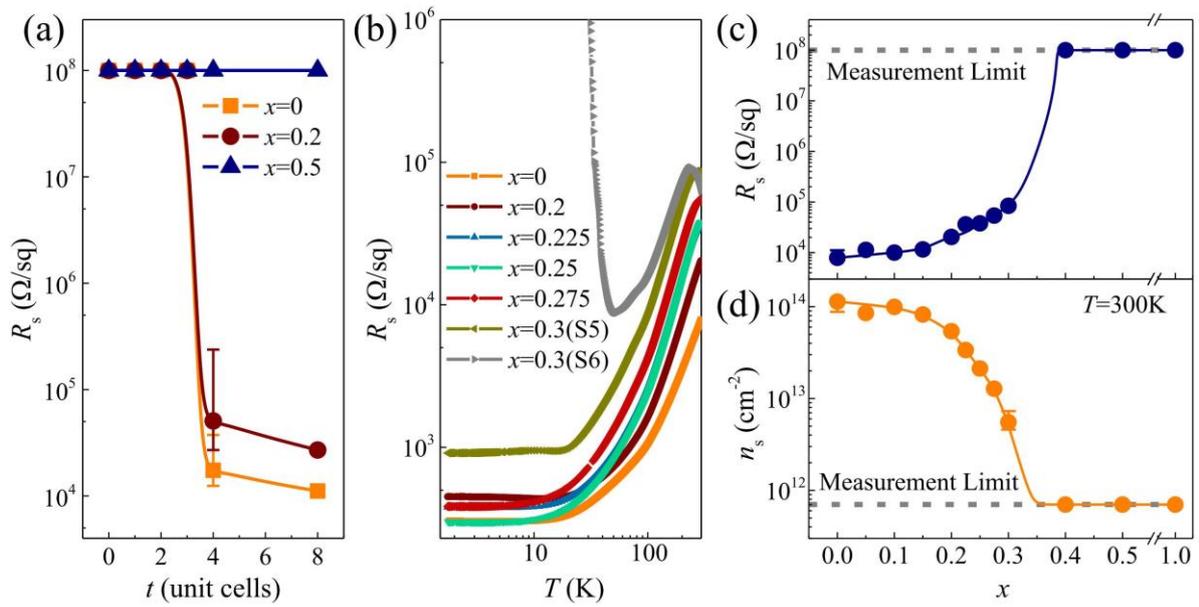

**Figure 3.** Mn-doping modulation of the transport properties at LAMO/STO interface ($0 \leq x \leq 1$). a) The critical thickness of LAMO films for interface conduction at $x = 0$, 0.2 and 0.5. b) Temperature dependence of the sheet resistance ($R_s$) at different Mn doping concentrations ($x$). c) and d) show the $R_s$ and carrier density ($n_s$), respectively, as a function of $x$, measured at room temperature. The change in the sheet resistance as a function of $x$ at room temperature is primarily caused by a change in the carrier density rather than the mobility. The solid lines are guides to the eye.

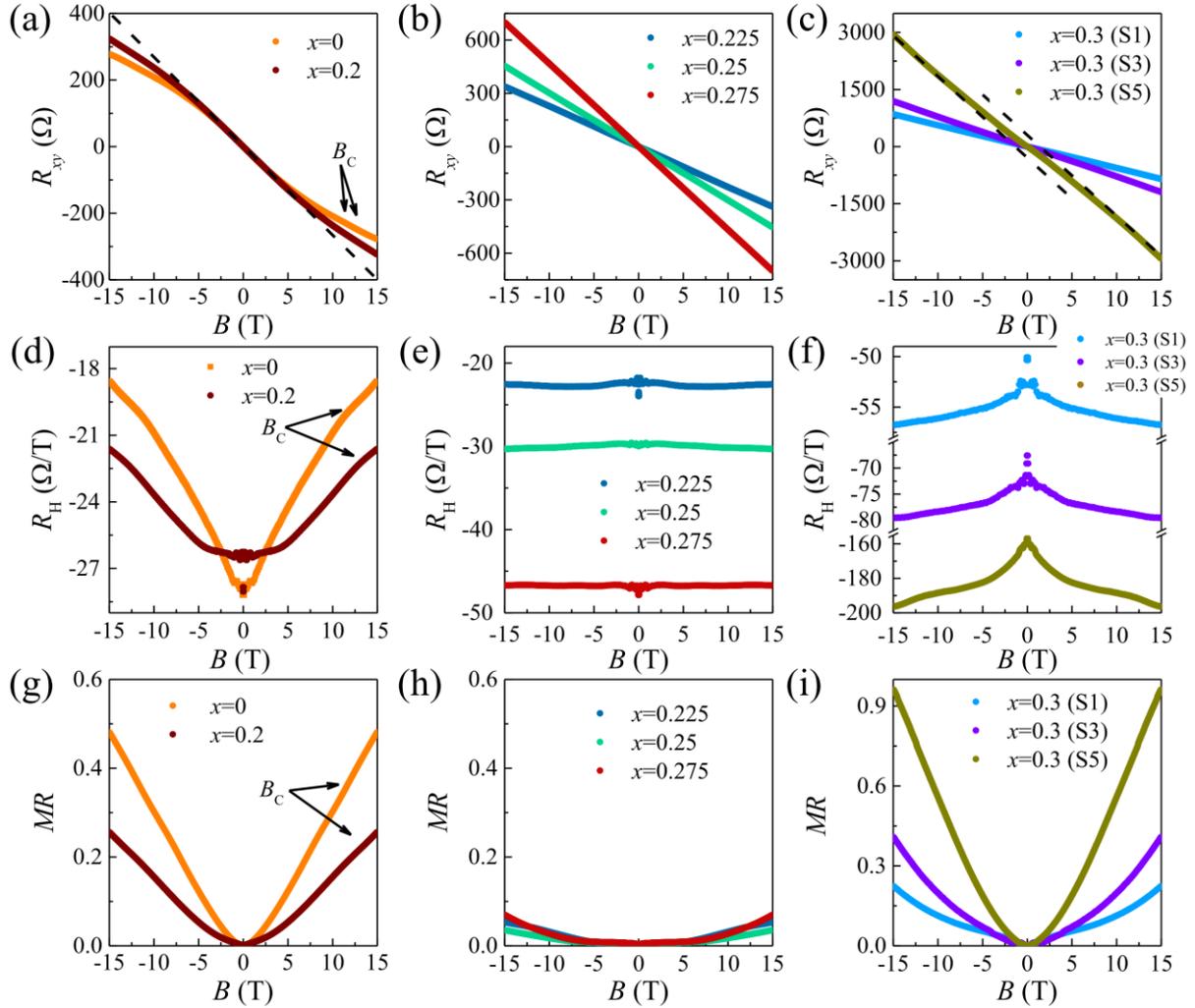

**Figure 4.** The evolution of transport properties of metallic LAMO/STO ($0 \leq x \leq 0.3$). a) Nonlinear Hall effect ($R_{xy}$) arising from two or more types of carriers for $x \leq 0.2$; b) Linear Hall effect originating from a single type of carriers for $0.225 \leq x \leq 0.275$; c) The anomalous Hall effect resulting from spin-polarized $d_{xy}$ electrons at $x = 0.3$. d)-f) The corresponding evolution of $R_H = R_{xy}/B$ calculated from (a-c). g) The bell-like MR curves for $x \leq 0.2$; h) The U-shaped MR curves for $0.225 \leq x \leq 0.275$; i) The U-shaped MR for $x = 0.3$. At $x = 0.3$, slight variations in carrier density change significantly the ground states, but all five metallic samples prepared at the same conditions show AHE behavior. They are marked from S1 to S5 with decreased carrier densities. All measurements were performed at 2 K.

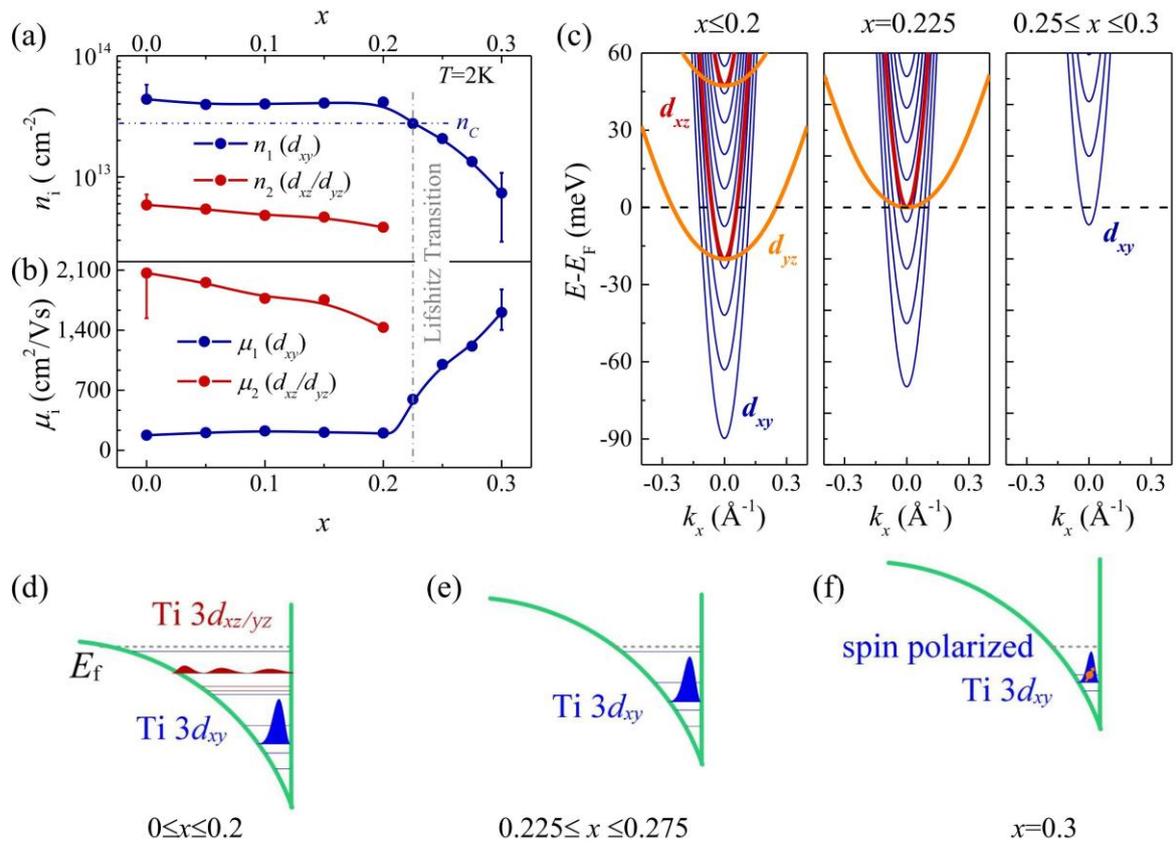

**Figure 5.** The evolution of electronic structure of LAMO/STO ($0 \leq x \leq 0.3$). Carrier density a) and mobility b) as a function of $x$ at 2 K. At $x \leq 0.2$, the values are extracted from a two-band model (The solid lines are guides to the eye). c) The band structures as a function of $x$ calculated by density functional theory based tight binding models. d)-f), Illustrations on the corresponding populations of orbitals and electronic states in Ti $t_{2g}$ bands for $x \leq 0.2$, $0.225 \leq x \leq 0.275$, and $x = 0.3$, respectively.

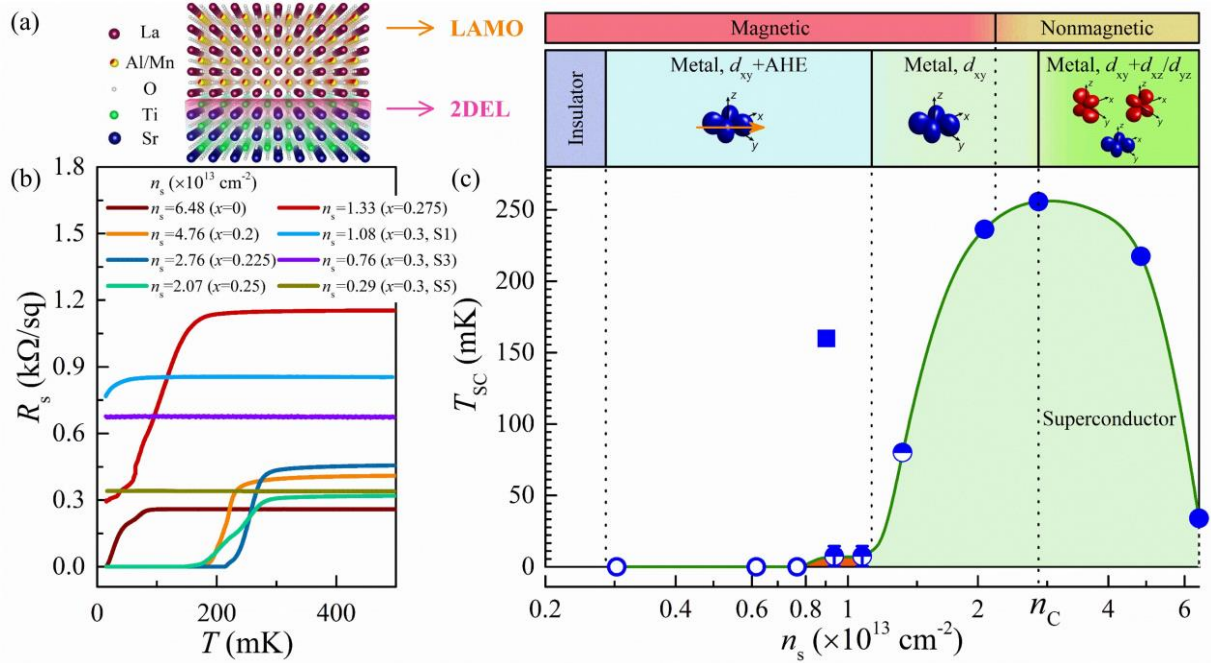

**Figure 6.** Phase diagram of LAMO/STO heterostructures. a) Sketch of the LAMO/STO heterointerface. b) Measured sheet resistance as a function of temperature down to 15 mK for typical samples. Measurements of all devices with different carrier densities are shown in Figure S10 in the supplementary information. c) The rich phase diagram of LAMO/STO heterostructures as a function of carrier density. At $x = 0$ (LAO/STO), the samples are metallic conducting with itinerant electrons occupying both $d_{xy}$ and $d_{xz}/d_{yz}$ bands and becoming superconducting below 34 mK. Here, $n_s$ denotes the total carrier density measured at 2 K. At a lower carrier density of $n_c = 2.8 \times 10^{13}$ cm$^{-2}$ ($x = 0.225$), the superconducting transition temperatures ($T_{SC}$) peaks at a value of 255 mK and here we observe a Liftshitz transition point. Below $n_c$, the $T_{SC}$ decreases and transport occurs only in the $d_{xy}$ band until the samples turn insulating ($n_s < 2.9 \times 10^{12}$ cm$^{-2}$). For $2.9 \times 10^{12}$ cm$^{-2}$ $\leq n_s \leq 1.1 \times 10^{13}$ cm$^{-2}$, the interface shows an anomalous Hall effect with occupation of only the $d_{xy}$ band. In addition, we observe a magnetically ordered state in the LAMO top film at $T \leq 21$ K for $x \geq 0.25$, whereas no signature of magnetism is observed below this doping level. Filled and open circles are used for superconducting and non-superconducting devices, respectively. Half-filled circles mark devices where the superconducting transition has initiated but do not reach a zero-resistant state at the base temperature of 15 mK. The solid line is a guide to the eye. Transport measurements are conducted on Hall bars whereas SQUID magnetometer measurements use unpatterned samples. Finally, the sample in a van der Pauw geometry ($n_s = 8.9 \times 10^{12}$ cm$^{-2}$) with full superconducting transition where $R_s = 0$ is reached at $T_{SC} \sim 160$ mK that also shows AHE is marked as filled square in the figure.